# Quasi-resonant neutralization of He$^+$ ions at a germanium surface

D Goebl[1,*], D Roth[1], D Primetzhofer[2], R C Monreal[3,4], E Abad[5], A Putz[1], and P Bauer[1]

[1] Institut für Experimentalphysik, Abteilung für Atom- und Oberflächenphysik, Johannes Kepler University Linz, 4040 Linz, Austria

[2] Institutionen för Fysik och Astronomi, Uppsala Universitet, Box 516, S-751 20 Uppsala, Sweden

[3] Departamento de Física Teórica de la Materia Condensada C5, Universidad Autónoma de Madrid, E-28049 Madrid, Spain

[4] Condensed Matter Physics Center (IFIMAC), Universidad Autónoma de Madrid, E-28049 Madrid, Spain

[5] Computational Biochemistry Group, Institute of Theoretical Chemistry, University of Stuttgart, D-70569 Stuttgart, Germany

E-mail: dominik.goebl@jku.at

**Abstract.** When low-energy He ions are scattered from a Ge surface, the fraction of positive ions exhibits characteristic oscillations as a function of ion energy. These oscillations are caused by quasi-resonant neutralization (qRN), a process which is active for materials with a narrow band nearly resonant with the unperturbed He-1s level. In the present manuscript we measure the fraction of He$^+$ backscattered from Ge(100). In conjunction with recently developed theoretical methods, we extract quantitative information on the efficiency of qRN. Our evaluation reveals that qRN is a highly efficient process leading to ion fractions two orders of magnitude lower than in systems for which neutralization is only due to Auger processes.

---

[*] Autor to whom correspondence should be addressed

Quasi-resonant neutralization of He$^+$ ions at a germanium surface

### I. Introduction

Charge exchange processes between ions and surfaces do play an important role in many fields throughout physics and chemistry, such as catalysis, plasma wall interactions, solar wind or nano-analytics. The present investigation is concerned with charge exchange in low-energy ion scattering (LEIS) [1, 2]. The basic idea of LEIS is to bombard a sample of interest with noble gas ions with a primary energy in the range of ~200 eV - 10 keV. Backscattered projectiles are detected by means of either an electrostatic analyzer (ESA-LEIS) or systems which measure the time-of-flight of the projectiles (TOF-LEIS). Based on the scattering geometry of the setup and the energy distribution of backscattered projectiles it is possible to determine surface composition and/or surface structure of the sample, respectively. As a surface analysis technique, LEIS is used in various fields throughout applied and basic research [3, 4, 5, 6]. Additionally, this method is an excellent tool to investigate charge exchange of slow noble gas ions with solid surfaces [7, 8].

The state-of-the-art description of charge exchange in LEIS distinguishes between five different processes: Auger-neutralization (AN), Auger ionization (AI), resonant neutralization (RN) and ionization (RI) in a close collision, and quasi-resonant neutralization (qRN). These processes lead to either neutralization of the impinging ion (AN, RN, qRN) or ionization of an already neutralized projectile (AI, RI). Other processes like resonant neutralization to excited states or double excitation to auto-ionizing states are omitted, since they do not play a considerable role for the He-Ge system. An important quantity which is experimentally accessible and related to charge exchange in LEIS is the fraction of positive ions $P^+$ amongst the backscattered projectiles.

Each of the charge exchange processes mentioned above may be the dominant mechanism, depending on the electronic properties of the surface and the distance of the ion to the surface. AN is possible whenever the unoccupied projectile-level is below the Fermi energy of the target. If this is the case, electrons can be transferred from a target state to the projectile level. The corresponding gain in binding energy is dissipated via excitation of either another electron or of a plasmon. The efficiency of AN usually is described by the AN transition-rate $\Gamma_A$. Since AN is a one-way process (neutralization only), $P^+$ plays the role of a surviving probability and is deduced from the rate equation $dP^+ = -P^+ \cdot \Gamma_A \cdot dt$:

$$P_i^+ = e^{-\int \Gamma_A(\vec{r}(t))dt}$$
$$= e^{-\int \Gamma_A \frac{dt}{dz}dz} \approx e^{-\frac{1}{v_z}\int \Gamma_A dz} = e^{-\frac{v_c}{v_{z,i}}} \quad (1)$$

Here, $i$ stands for the inbound or outbound part of the trajectory $\vec{r}(t)$. The characteristic velocity $v_c = \int \Gamma_A \cdot dz$ measures the AN efficiency of a given projectile-target combination. The integral boundaries are chosen corresponding to start and end points of the respective part of the trajectory ("in" or "out"). In absence of other charge exchange processes, i.e., when neutralization is exclusively due to AN, $P^+$ can be written as:

$$P^+ = P_{in}^+ \cdot P_{out}^+ \approx e^{-v_c(\frac{1}{v_{z,in}} + \frac{1}{v_{z,out}})} \equiv e^{-\frac{v_c}{v_\perp}} \quad (2)$$

where $v_\perp$ denotes the inverse perpendicular velocity.

When the ion approaches the surface, the interaction between He$^+$ and the surface leads to a shift of the unoccupied projectile level, i.e. the ionization potential, due to various processes: At an ion-surface distance of ~5 atomic units (a.u.), image charge interaction leads to a noticeable promotion of the level [9]. When the projectile approaches further, the interaction with the conduction band electrons may lead to an increase in binding energy of the level. At even closer distances the interaction with core level electrons of the target leads to a strong upward shift of the He 1$s$-level [10, 11]. If at some small distance the unoccupied level becomes resonant with the conduction band, the projectile can be neutralized by resonant charge exchange. An even stronger shift will eventually lead to a promotion of the level above $E_F$, thus enabling resonant ionization of already neutralized projectiles. These processes require the projectile to approach a target atom to a distance smaller than a certain minimum distance, $r_{min}$. Since scattering angles of the different LEIS setups are very similar, this minimum distance is often substituted by a more practical threshold energy for collision induced processes $E_{th}$.

RN, RI and AN are charge exchange processes which are not limited to certain target-projectile combinations. In recent years, many experimental and theoretical studies were devoted to gaining a better understanding of these processes [9, 12, 13, 14, 15, 16, 17]. In addition to the processes discussed so far, for certain projectile-target combinations an additional charge exchange

Quasi-resonant neutralization of He$^+$ ions at a germanium surface mechanism is possible: quasi-resonant neutralization (qRN). Although this process has been known for many years, quantitative information is not available up to now. To shed light on the efficiency of qRN relative to AN and the resonant charge transfer processes in a close collision, the present investigation focuses on a quantitative analysis of this process for the He-Ge system. Quantitative data is the key to validate theoretical predictions for the different neutralization processes.

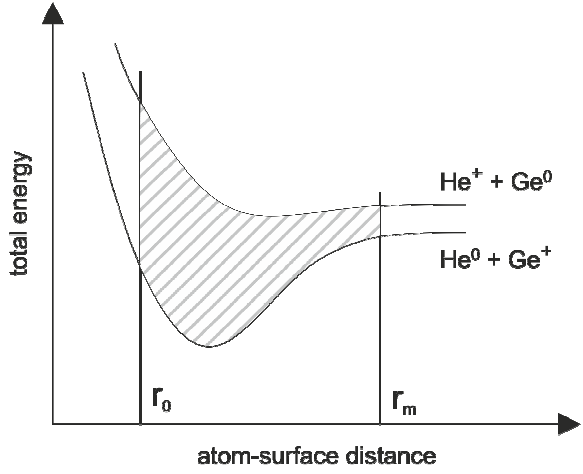

*Figure 1. Energy diagram of qRN in the He-Ge system. Here, $r_m$ denotes the mixing distance, and $r_0$ indicates the turning point of the trajectory.*

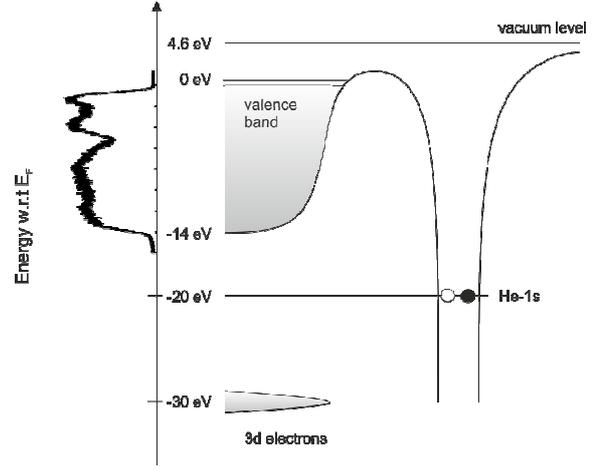

*Figure 2. Sketch of the electronic configuration of the He-Ge system. Corresponding energies are taken from [27, 28]; the displayed DOS is taken from [28].*

## II. Quasi-resonant charge exchange

The first evidence for qRN of He$^+$ ions in LEIS was reported by Erickson and Smith [18]. They observed oscillations of the ion yield as a function of primary energy for various target materials: Ge, In, Pb and Bi. A first semi-classical explanation for these oscillations was published soon after by Tolk et al. [19], which traced the oscillations in the ion yield back to a quantum interference effect. When a He$^+$ ion approaches a surface atom X, quasi resonant charge exchange can occur at inter-nuclear distances $r$ smaller than a certain "mixing distance" $r_m$. At $r < r_m$ a coherent mixed state He$^+$ + X and He$^0$ + X$^+$ evolves. For the time the ion spends at a distance smaller than $r_m$ a phase difference $\Delta\varphi$ develops which depends on the difference between the specific potential curves ($\Delta E$) of the different states (see Fig. 1). The total phase difference can be calculated by evaluating the following equation:

$$\Delta\varphi = \frac{1}{\hbar}\int \Delta E(t)dt = \frac{1}{\hbar}\int \frac{\Delta E(\vec{R})}{\bar{v}(\vec{R})}dR \qquad (3)$$

The integral boundaries correspond to the points when the projectile passes $r_m$ on the inbound and outbound part of its trajectory; $\bar{v}$ denotes the mean projectile velocity. Note, that qRN is an atomic process, which implies that - contrary to AN - qRN depends on the inverse *initial velocity* $v_0$ instead of the inverse perpendicular velocity $v_\perp$. From the total phase difference, the intensity of backscattered ions $I^+$ can be written as $I^+ = A^+ + B^+ \sin^2(\Delta\varphi/2)$, where the coefficients $A^+$, and $B^+$ are slowly-varying functions of the ion energy. Already Tolk's approach leads to an at least qualitative agreement with experimental data. Since then, various efforts have been dedicated to obtaining a more detailed understanding of qRN [20, 21, 22, 23, 24, 25].

So far, oscillations in the ion yield have only been observed for target materials which exhibit either *d*- or *f*-band electrons with a binding energy nearly resonant with the ionization potential of He$^+$ [1,26].

In the present investigation, qRN of He$^+$ at a Ge(100) surface is studied. A sketch of the electronic configuration of the He-Ge system is given in Fig. 2, assuming a work function of 4.6 eV, in good agreement with [27]. As one can see, Ge exhibits a very broad valence band which is filled by *p*- and *s*-electrons. The DOS of the valence band is also shown in Fig. 2. Due to the lack of *d*-electrons, it does not exhibit pronounced features. The 10 *d*-electrons form a very sharp band at ~30 eV below $E_F$ [28].

## III. Sample preparation, experiment and evaluation

Experiments have been conducted using a p-type Ge(100) sample in the TOF-LEIS setup ACOLISSA [29] and the ESA-LEIS setup MINIMOBIS [30]. In order to obtain a clean surface, multiple sputtering-annealing cycles were performed, using 3 keV Ar$^+$ ions and annealing temperatures of ~600 °C. Surface cleanness was checked by Auger-Electron-Spectroscopy (ACOLISSA) and by LEIS (MINIMOBIS, ACOLISSA). Low energy electron diffraction (LEED) was employed to examine the crystalline quality of the surface. LEED images were recorded and exhibited sharp spots and a clear (2x1) structure with 2 domains rotated by 90° [31].

From spectra recorded with an ESA-LEIS setup, $P^+$ is determined from the ion yield $Y^+$ as follows:

$$P^+ = \frac{c \cdot Y^+}{I \cdot t \cdot n \cdot \frac{d\sigma}{d\Omega} \cdot \eta^+} \quad (4)$$

Here, $I$ denotes the primary beam current, $t$ is the acquisition time per data point, $n$ stands for the surface areal density, $d\sigma/d\Omega$ denotes the scattering cross section, $\eta^+$ represents the detection efficiency. Other experimental quantities like the detector solid angle and spectrometer transmission are included in the experimental factor $c$. The scattering cross section is calculated based on the Thomas-Fermi-Moliere screened potential [32] with the Firsov screening length [33] corrected by a factor of 0.75, which had been found adequate for Cu [34]. Since our spectrometer features a $2\pi$ azimuthal acceptance angle, it is a non-trivial task to define the exact surface areal density of Ge (100). Thus, $P^+$ values obtained using our ESA-LEIS setup were normalized to match the TOF-LEIS results. Due to the different azimuthal acceptance in the two setups, one might expect a systematically different energy dependence of the deduced $P^+$ data. The fact, however, that both data sets coincide perfectly within statistics gives strong evidence for the very minor subsurface contributions to the ion yield (see Fig. 3). Note, that due to the normalization of the ESA-LEIS data, $P^+$ is independent of possible uncertainties in the scattering potential and the effective surface areal density.

Determination of $P^+$ from TOF-LEIS spectra is straightforward, since this type of setup detects both, ions and neutrals. To separate ions from neutrals our system makes use of a post-acceleration lens located between sample and detector. For certain crystals, in double alignment geometry the information depth can be limited to the outermost atomic layer. In this case, $P^+$ can be evaluated directly from the yield of detected ions, $Y^+$, and of detected neutrals, $Y^0$:

$$P^+ = \frac{Y^+}{Y^+ + Y^0 \cdot \eta^+ / \eta^0} \quad (5)$$

For $Y^0$ only the single scattering peak is relevant. Therefore, contributions from deeper layers are eliminated by proper background subtraction. With this method, $P^+$ values can be determined without exact knowledge of scattering cross section, detector solid angle and surface areal density. The detector efficiencies for neutrals, $\eta^0$, are determined from reference measurements with Cu samples; $\eta^+$ is close to 1 since the first MCP is at a potential of -2.4 kV.

For the Ge(100) (2×1) surface, it is expected that in double alignment geometry always two monolayers are contributing to the yield of neutrals, $Y^0$. To the ion yield, $Y^+$, these two monolayers contribute differently, due to the fact that for the projectiles backscattered from the second layer the probability to avoid Auger neutralization is lower. Therefore, the ion fraction obtained from Eq. (5) represents an *apparent* ion fraction slightly lower than expected for only one atomic layer contributing. Nevertheless, as argued above, the ion yield from the outermost surface will by far exceed that from subsurface layers.

Polar scans are a sensitive method to obtain information on charge exchange by RI. In a polar scan, the sample is tilted along the polar axis (perpendicular to the scattering plane). The value of $P^+$ for a certain polar angle $\alpha$ can then be derived from a reference $P_0^+$, recorded in double alignment geometry:

$$P^+(\alpha) = P_0^+ \frac{Y^+(\alpha)}{Y_0^+} \cdot \cos(\alpha) \quad (6)$$

In our case normal incidence ($\alpha = 0°$) was chosen as the reference angle. When the polar angle is changed, a substantial increase in the yield of backscattered projectiles is observed, as the number of visible layers rises. The number of visible atoms in the topmost layer will increase by a factor $1/\cos(\alpha)$. If RI processes occur, projectiles which are backscattered from deeper layers have a chance to be re-ionized on their way out of the sample. As a consequence, these projectiles lead to an increase of $Y^+$ when double alignment conditions are abandoned [35]. To minimize systematic errors, a polar scan requires a constant primary beam current. Therefore, the beam current was checked in between measurements for different polar angles.

## IV. Results and Discussion

Fig. 3 presents $P^+$ values measured for He$^+$ scattered from Ge(100) as a function of inverse initial velocity, $1/v_0$. Data were deduced from measurements taken in our TOF- (black squares) and ESA-LEIS (red circles) systems. One can clearly observe the prominent oscillations in $P^+$ in the energy range of 575 eV ($1/v_0 = 6.0 \cdot 10^{-6}$ s/m) to 5.2 keV ($1/v_0 = 2.0 \cdot 10^{-6}$ s/m). At very low energies, the oscillations are damped and a resolution of individual minima and maxima is not possible. The oscillation period amounts to $6.7 \cdot 10^{-7}$ s/m (±10 %). Note that the absolute $P^+$ values are very low compared to ion-target combinations where AN prevails. To illustrate this fact, $P^+$ values for He$^+$ scattered from polycrystalline Cu are included in Fig. 3 (open squares) [36]. These values are almost one order of magnitude higher than $P^+$ values for Ge(100). This makes clear that for the He-Ge system, AN plays a minor role and qRN is the dominant neutralization process. Therefore, it is appropriate to present these data as a function of $1/v_0$ rather than $1/v_\perp$.

In Fig. 4, our experimental values from ESA-LEIS are compared to the existing data obtained by Erickson and Smith [18]. In [18], the results were presented as ion yield in arbitrary units. To allow a direct comparison of these data to our $P^+$ results, it was necessary to correct the data from [18] for the energy dependence of the scattering cross section. Normalization was performed such that $P^+$ coincides with our data at the oscillation maximum at ~1 keV. Note that although the setup of Erickson and Smith employs a scattering angle (90°) significantly different from our scattering angle (136°), maxima and minima in the oscillations coincide remarkably well. This confirms the finding of reference [19] that for sufficiently large scattering angles and sufficiently high projectile energies the maxima do not significantly depend on the scattering angle. Due to the conversion procedure to deduce $P^+$ from the ion yield data in [18], comparison of the absolute values is of limited value. Nevertheless, significant deviations appear only for energies above 1.5 keV.

We now want to discuss the relative importance of the different charge exchange processes, i.e., AN, RN, RI and qRN. As a first step we determine the threshold energy, $E_{th}$, for the resonant processes RI and RN. To this end, ESA-

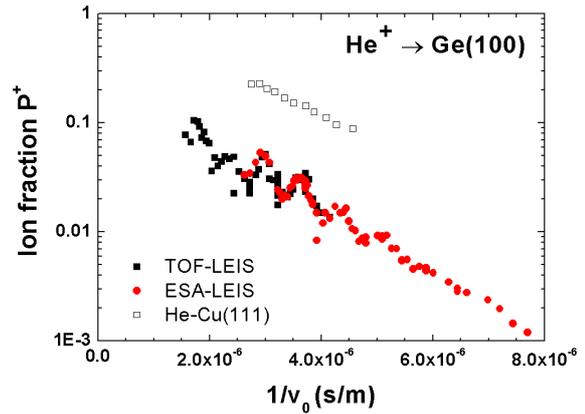

*Figure 3.* $P^+$ for He$^+$ scattered from Ge(100). Data were obtained from measurements in our TOF-LEIS (black squares) and ESA-LEIS (red circles) setups. To illustrate the efficiency of qRN, $P^+$ values are shown for He$^+$ scattered from polycrystalline Cu (open squares), where neutralization is exclusively due to AN [36].

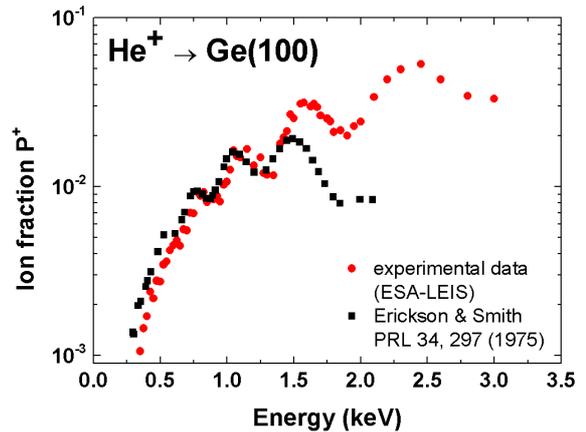

*Figure 4.* $P^+$ values measured in our ESA-LEIS system compared to existing data measured by Erickson and Smith [18].

LEIS spectra were recorded for an incident energy of 4 keV. The obtained spectrum, which is presented in Fig. 5, exhibits a single scattering peak at ~3200 eV followed by a tail which can be interpreted in the following way: projectiles are neutralized when penetrating the sample, they are backscattered from deeper layers, experience RI in a final collision close to the surface and remain in the positive charge state when leaving the surface. The final energy $E_f$ of the projectile depends on the path length travelled in the sample; when $E_f$ falls below $E_{th}$, RI does not contribute to the ion yield anymore. There, the observed background intensity is only due to noise and to secondary ions with very low kinetic energy. Thus, it is possible to deduce $E_{th}$ from the onset energy of the tail, $E_T \approx 1100$ eV. This onset corresponds to $E_f$, while $E_{th}$ is usually referred to the incident energy, $E_{th} = E_T/k \approx 1300$ eV, with $k$ denoting the kinematic factor [37].

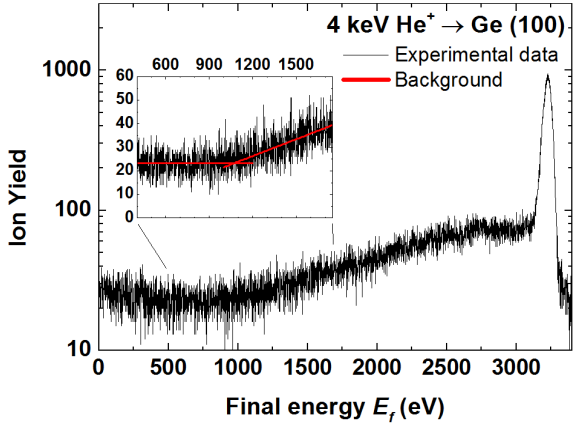

*Figure 5.* ESA-LEIS spectrum of 4 keV He+ scattered from a Ge surface. In the inset, one can see the part of the spectrum where the tail of the binary collision peak starts to raise from the background noise.

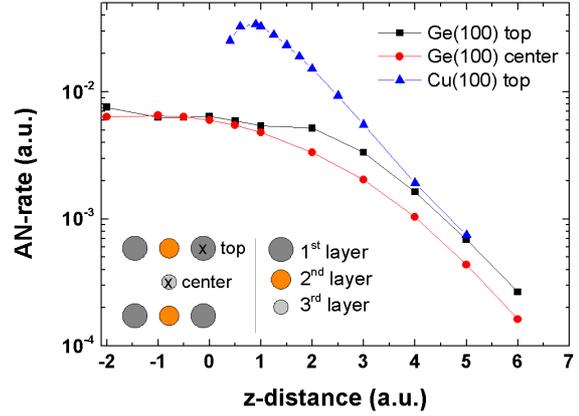

*Figure 6.* AN-rates for He+ in front of a Ge(100) surface as a function of atom-surface distance. Rates were calculated for the position indicated in the inset. The results for Ge are compared to previously obtained rates for Cu(100) [16]. Distance z is measured with respect to the topmost atomic layer.

As a next step, the relative importance of AN and qRN is analyzed at energies below $E_{th}$. In order to quantify the contribution due to AN, we performed calculations using a model based on a LCAO description of the target electrons [15]. This model has already been successfully applied to investigate AN in the case of the noble metal surfaces [16], where it was shown that the AN-rate is sensitive to the exact position of the He $1s$-level. However, this effect is less pronounced for materials where $d$-electrons are absent, e.g. for Al [38, 39], or do not play an active role in the neutralization process, as for Ge (see Fig. 2). AN-rates were calculated for the unreconstructed Ge(100) surface. The density of states (DOS), which is necessary for the calculation, was approximated to fit results from density functional theory (DFT) calculations [40]. The resulting AN-rates for the on-top position (squares) and the center position of the Ge(100) unit cell are shown in Fig. 6. To allow for a comparison with other materials, results obtained for Cu(100) are also shown in the plot. At large ion-surface distances, i.e., above 4 a.u. Cu and Ge exhibit similar AN-rates, but at a distance of ~1 a.u. Cu rates exceed Ge rates by more than a factor of 5, due to the efficient contribution of the Cu-$3d$ electrons [16].

From Eq. (1), the survival probability in the presence of AN can be calculated along the trajectories $\vec{r}(t)$ employing the theoretical AN-rates $\Gamma_A(\vec{r}(t))$. The trajectory simulations were performed by means of the MD simulation program KALYPSO [41] using the ZBL potential [42] without a screening length correction. The surface reconstruction was not implemented in the simulations, however, two different azimuthal orientations (0° and 90°) were employed to model the 2-domain structure of the surface as observed in LEED. Depending on the azimuthal angle, projectiles backscattered from either the first and second layers (90°) or the first and third layers (0°) contribute to the ion yield. From these simulations, $P^+$ due to AN-only is obtained by averaging. The results (full circles) are shown in Fig. 7 together with the experimental data (open squares).

For all investigated incident energies, AN is only a minor contribution to neutralization. For energies lower than $E_{th}$, qRN dominates and leads to $P^+$ values which are ~1-2 orders of magnitude lower than anticipated for AN only. The uncertainty in the calculated AN-rates due to the applied approximations is far too small to explain such large discrepancies. In previous investigations [21, 23], it was argued that only one neutralization channel is active at a time; consequently AN and qRN were treated separately. The z-dependence of the obtained AN-rates does not a-priori justify this approximation. However, since the probability for AN is very low, the decoupling of the different neutralization channels, $P^+ = P^+_{AN} \cdot P^+_{qRN}$, does not lead to a substantial error. Single exponential fits to the AN data and to the experimental data at low energies yield characteristic velocities that differ by a factor of ~ 5 A similar value may be expected for the ratio $(\Gamma_{AN}+\Gamma_{qRN})/\Gamma_{AN}$, assuming the two processes to be active within equal time

Quasi-resonant neutralization of He⁺ ions at a germanium surface windows. In a very simple description of $P^+$ in the investigated energy range, all charge exchange processes are treated as independent of each other. Consequently, $P^+$ can be written as a product of probabilities: $P^+ = P^+_{AN} \cdot P^+_{qRN} \cdot P^+_{RN}$. $P^+_{AN}$ is described by a single exponential: $P^+_{AN} = exp(-v_{AN}/v_0)$, with $v_{AN} = 1.7 \cdot 10^5$ m/s (see Fig. 7). The term for $P^+_{qRN}$ contains a single exponential and an oscillatory term: $P^+_{qRN} = exp(-v_{qRN}/v_0) \cdot [1+ a^+(E) \cdot sin(\pi \cdot v_{osc}/v_0 - \Delta\phi)^2]$, with $v_{qRN} = 7.4 \cdot 10^5$ m/s. For the energy dependent amplitude of the oscillations, $a^+(E)$ a Gaussian shape was assumed on the basis that the largest amplitude is observed at ~3 keV; at lower and higher energies damping maybe expected due to AN and due to resonant processes, respectively. The oscillation velocity, $v_{osc} \approx 1.41 \cdot 10^6$ m/s and the phase shift $\Delta\phi \approx -0.9$ characterize the observed oscillations, in good quantitative agreement with earlier experiments [25]. The influence of resonant charge exchange in close collisions, $P^+_{RN,}$ is globally taken into account by lowering $P^+$ at energies above $E_{th}$ by use of a Fermi-Dirac distribution like function. This description leads to good agreement with experiment, as shown in Fig. 7 (solid line)

From this simple model hardly any detailed information for the individual processes can be deduced. A more advanced theory should include the following key quantities [21, 23, 25]:

(1) The mismatch in energy between the He ground state and the Ge $d$-band

(2) The distance-dependent level-shift of the He ground state

(3) The width of the $d$-band

(4) Resonant tunneling to an excited state [23] is not deemed a factor here due to the high work-function of Ge.

Points (1) and (2) make clear that – as for AN – the exact knowledge of the distance- dependence of the He level is crucial to gain more insight into the qRN process. This might also explain why the electron transfer is so efficient for very low energies: while at large distance the mismatch between the He ground state and the Ge-3$d$ level is pronounced, it may be significantly lower at the distance where qRN occurs. To analyze the interplay of the neutralization channels AN, RN, RI, and qRN thoroughly, advances in the theoretical description are required.

Finally, we deduce information on the role of reionization from polar scans, taken at 1.8 and 2.3 keV (see Fig. 8). These energies are already in the reionization regime and correspond to a minimum and a maximum, respectively, in the $P^+$ oscillations. As can be seen, both polar scans reveal either an almost constant or an unremarkable dependence of $P^+$ on the polar angle. In other words, the yield of backscattered ions does not change when double alignment conditions are abandoned (apart from the trivial $1/cos\alpha$ dependence, see. Eq. 6). Significant resonant reionization would result in a breakdown of the $1/v_\perp$- scaling (see below). Since this is not the case, we conclude that at least for energies up to 2.3 keV RI does not play a significant role. To estimate the contribution of RI at incident energies which are far in the reionization regime, a polar scan at 6 keV was performed (see Fig. 9). This scan reveals pronounced features in the $\alpha$-dependence of $P^+$: when the channeling condition is abandoned, the apparent $P^+$ increases by a factor of ~2 due to an increase of the depth from which reionized projectiles can survive in the positive charge state. Whenever the probability for RI is considerable, an even higher efficiency for RN has to be expected [17]. Nevertheless, since the mixing distance is expected to be considerably larger [19] than the minimum distance for collision induced resonant processes, $r_{min}$, quasi-resonant neutralization does not need a large angle scattering event to be possible. This may explain why for Ge also in the reionization regime qRN is the dominant neutralization mechanism - $P^+$ still is very low and exhibits significant oscillations as one can see in Fig. 3.

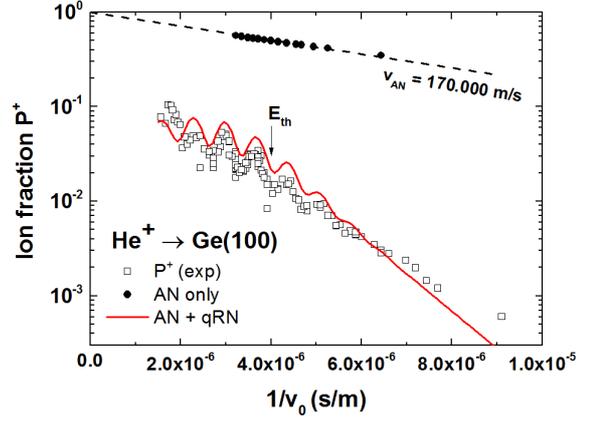

*Figure 7. Ion fraction of He⁺ backscattered from Ge(100). Experimental results are indicated by open symbols. The calculated $P^+$ data according to AN-only (full circles) is fitted by a single-exponential (dashed line). Additionally, results of a simple model which includes all charge exchange process are shown (red solid line).*

Quasi-resonant neutralization of He+ ions at a germanium surface

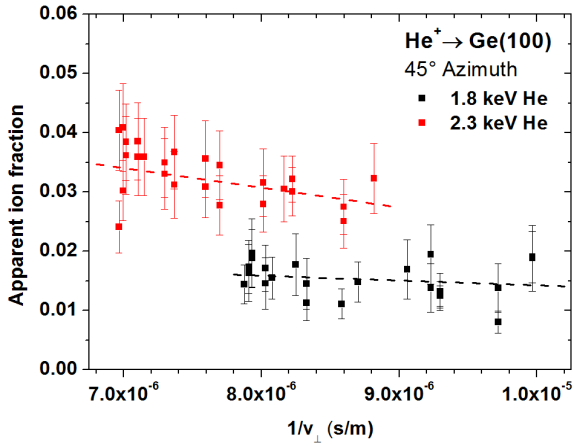

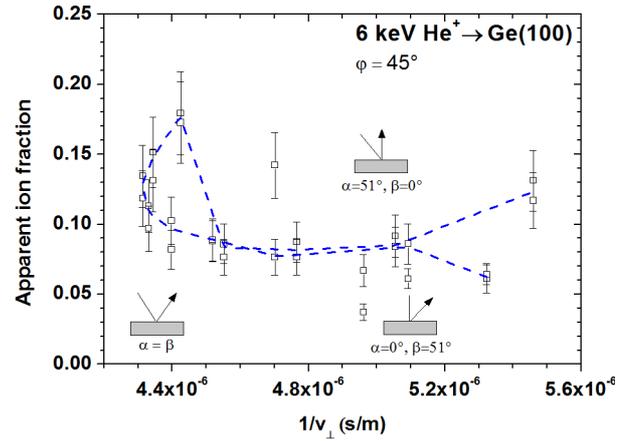

*Figure 8.* Polar scan of He$^+$ at Ge(100) at 1.8 keV and 2.3 keV (the energies correspond to a minimum or maximum, respectively, of the $P^+$ oscillations (see Fig. 3). $P^+$ exhibits only minor variation which indicates that RI does not play a significant role at these energies.

*Figure 9.* Polar scan of He$^+$ scattered from Ge(100) at 6 keV. $P^+$ exhibits a significant increase when abandoning channeling geometry. Note that the error bars in this plot are only based on counting-statistics. Other factors, such as beam stability, were not considered when calculating the error bars, but can explain single discordant values.

## V. Summary and outlook

We have obtained quantitative information on the ion fraction $P^+$ for He$^+$ scattered from a Ge(100) surface in the energy range of 350 eV – 8.5 keV. Strong oscillations of $P^+$ as a function of initial energy were observed for an energy range from 575 eV to 5.2 keV. Obtained data are lower by almost one order of magnitude compared to $P^+$ data for materials where neutralization is exclusively due to AN. The threshold energy for collision induced charge exchange was determined to be ~1.3 keV. However, polar scans have shown that the probability for reionization in close collisions is small - at least for energies below 2.3 keV. At higher energies, e.g., 6 keV, the reionization probability increases which manifests itself in a rise of the apparent $P^+$ by a factor of ~2 when channeling conditions are abandoned. At energies below the reionization threshold we have calculated the ion fractions resulting from the pure AN process and shown in this way that qRN is much more effective than AN; in this regime, a simple charge exchange model has been applied to reproduce the $P^+$ oscillations qualitatively.

For a better understanding of the qRN process, a more sophisticated theory would be highly desirable. This theory should model the distance-dependent level-shift of the He ground state in front of a Ge surface in a realistic way and determine $P^+$ when neutralization may be due to AN or qRN. A full description, however, should take all charge exchange processes into account. Since charge exchange is very sensitive to the band structure of the target material, such a description would be a highly sensitive tool to monitor changes in the band structure, e.g., when going to nanometer-sized clusters.

### Acknowledgement

Financial support by the Austrian Science Fund (FWF): project P22578 is acknowledged. D. Goebl acknowledges support as a DOC-Fellow of the Austrian Academy of Sciences. R.C. Monreal acknowledges financial support by the Spanish MINECO, project FIS2011-26516